\documentstyle[twoside]{article}

\catcode`\@=11
\long\def\@makefntext#1{ %\parindent 1em
\protect\noindent \hbox to 3.2pt {\hskip-.9pt
$^{{\eightrm\@thefnmark}}$\hfil}#1\hfill} %can be used

\def\thefootnote{\fnsymbol{footnote}}
 \def\@makefnmark{\hbox to 0pt{$^{\@thefnmark}$\hss}}  %original

\def\ps@myheadings{\let\@mkboth\@gobbletwo
\def\@oddhead{\hbox{} %\sl
\rightmark\hfil\eightrm\thepage}
\def\@oddfoot{}\def\@evenhead{\eightrm\thepage\hfil %\sl
\leftmark\hbox{}}\def\@evenfoot{}
\def\sectionmark##1{}\def\subsectionmark##1{}}

\oddsidemargin=\evensidemargin
\renewcommand{\thefootnote}{\fnsymbol{footnote}}

\newcounter{sectionc}\newcounter{subsectionc}\newcounter{subsubsectionc}
\renewcommand{\section}[1] {\vspace{12pt}\addtocounter{sectionc}{1}
\setcounter{subsectionc}{0}\setcounter{subsubsectionc}{0}\noindent
        {\tenbf\thesectionc. #1}\par\vspace{5pt}}
\renewcommand{\subsection}[1] {\vspace{12pt}\addtocounter{subsectionc}{1}
        \setcounter{subsubsectionc}{0}\noindent
        {\bf\thesectionc.\thesubsectionc. {\kern1pt \bfit #1}}\par\vspace{5pt}}
\renewcommand{\subsubsection}[1] {\vspace{12pt}\addtocounter{subsubsectionc}{1}
        \noindent{\tenrm\thesectionc.\thesubsectionc.\thesubsubsectionc.
        {\kern1pt \tenit #1}}\par\vspace{5pt}}
\newcommand{\nonumsection}[1] {\vspace{12pt}\noindent{\tenbf #1}
        \par\vspace{5pt}}

\newcounter{appendixc}
\newcounter{subappendixc}[appendixc]
\newcounter{subsubappendixc}[subappendixc]
\renewcommand{\thesubappendixc}{\Alph{appendixc}.\arabic{subappendixc}}
\renewcommand{\thesubsubappendixc}
        {\Alph{appendixc}.\arabic{subappendixc}.\arabic{subsubappendixc}}

\renewcommand{\appendix}[1] {\vspace{12pt}
        \refstepcounter{appendixc}
        \setcounter{figure}{0}
        \setcounter{table}{0}
        \setcounter{lemma}{0}
        \setcounter{theorem}{0}
        \setcounter{corollary}{0}
        \setcounter{definition}{0}
        \setcounter{equation}{0}
        \renewcommand{\thefigure}{\Alph{appendixc}.\arabic{figure}}
        \renewcommand{\thetable}{\Alph{appendixc}.\arabic{table}}
        \renewcommand{\theappendixc}{\Alph{appendixc}}
        \renewcommand{\thelemma}{\Alph{appendixc}.\arabic{lemma}}
        \renewcommand{\thetheorem}{\Alph{appendixc}.\arabic{theorem}}
        \renewcommand{\thedefinition}{\Alph{appendixc}.\arabic{definition}}
        \renewcommand{\thecorollary}{\Alph{appendixc}.\arabic{corollary}}
        \renewcommand{\theequation}{\Alph{appendixc}.\arabic{equation}}
        \noindent{\tenbf Appendix \theappendixc #1}\par\vspace{5pt}}
\newcommand{\subappendix}[1] {\vspace{12pt}
        \refstepcounter{subappendixc}
        \noindent{\bf Appendix \thesubappendixc. {\kern1pt \bfit #1}}
        \par\vspace{5pt}}
\newcommand{\subsubappendix}[1] {\vspace{12pt}
        \refstepcounter{subsubappendixc}
        \noindent{\rm Appendix \thesubsubappendixc. {\kern1pt \tenit #1}}
        \par\vspace{5pt}}

\newcommand{\textlineskip}{\baselineskip=13pt}
\newcommand{\smalllineskip}{\baselineskip=10pt}

\def\eightcirc{
\begin{picture}(0,0)
\put(4.4,1.8){\circle{6.5}}
\end{picture}}
\def\eightcopyright{\eightcirc\kern2.7pt\hbox{\eightrm c}}

\newcommand{\copyrightheading}[1]
        {\vspace*{-2.5cm}\smalllineskip{\flushleft
        {\eightrm International Journal of Modern Physics D, #1}\\
        {\eightrm $\eightcopyright$\, World Scientific Publishing
         Company}\\
         }}

\newcommand{\publisher}[2]{{\begin{center}\eightrm\smalllineskip
        Received #1\\
        Revised #2
        \end{center}
        }}

\def\abstracts#1#2#3{{
        \centering{\begin{minipage}{4.5in}\baselineskip=10pt\eightrm
        \centerline{ABSTRACT}
        \parindent=0pt #1\par
        \parindent=15pt #2\par
        \parindent=15pt #3
        \end{minipage} }\par}}

\renewenvironment{thebibliography}[1]                   %ALL CHANGES DD 13/3/92
        {\ninerm
         \baselineskip=11pt                             %changed by cheng
         \begin{list}{\arabic{enumi}.}
        {\usecounter{enumi}\setlength{\parsep}{0pt}
         \setlength{\leftmargin 17pt}{\rightmargin 0pt} %changed by cheng
         \setlength{\itemsep}{0pt} \settowidth          %changed by cheng
        {\labelwidth}{#1.}\sloppy}}{\end{list}}

\newcounter{itemlistc}
\newcounter{romanlistc}
\newcounter{alphlistc}
\newcounter{arabiclistc}

\newcommand{\fcaption}[1]{
        \refstepcounter{figure}
        \setbox\@tempboxa = \hbox{\eightrm Fig.~\thefigure. #1}
        \ifdim \wd\@tempboxa > 5in
           {\begin{center}
        \parbox{5in}{\eightrm \smalllineskip Fig.~\thefigure. #1 }
            \end{center}}
        \else
             {\begin{center}
             {\eightrm Fig.~\thefigure. #1}
              \end{center}}
        \fi}

\newcommand{\tcaption}[1]{
        \refstepcounter{table}
        \setbox\@tempboxa = \hbox{\eightrm Table~\thetable. #1}
        \ifdim \wd\@tempboxa > 5in
           {\begin{center}
        \parbox{5in}{\eightrm\smalllineskip Table~\thetable. #1 }
            \end{center}}
        \else
             {\begin{center}
             {\eightrm Table~\thetable. #1}
              \end{center}}
        \fi}

\def\@citex[#1]#2{\if@filesw\immediate\write\@auxout    %IJMPA, IJMPB ONLY
        {\string\citation{#2}}\fi                       %TO DELETE PERCENTAGE
\def\@citea{}\@cite{\@for\@citeb:=#2\do                 %KEY WHEN USING
        {\@citea\def\@citea{,}\@ifundefined             %DD 13/3/92
        {b@\@citeb}{{\bf ?}\@warning
        {Citation `\@citeb' on page \thepage \space undefined}}
        {\csname b@\@citeb\endcsname}}}{#1}}

\newif\if@cghi
\def\cite{\@cghitrue\@ifnextchar [{\@tempswatrue
        \@citex}{\@tempswafalse\@citex[]}}
\def\citelow{\@cghifalse\@ifnextchar [{\@tempswatrue
        \@citex}{\@tempswafalse\@citex[]}}
\def\@cite#1#2{{$\null^{#1}$\if@tempswa\typeout
        {IJCGA warning: optional citation argument
        ignored: `#2'} \fi}}

\def\pmb#1{\setbox0=\hbox{#1}
        \kern-.025em\copy0\kern-\wd0
        \kern.05em\copy0\kern-\wd0
        \kern-.025em\raise.0433em\box0}

\def\fnt#1#2{\footnotetext{\kern-.3em
        {$^{\mbox{\scriptsize #1}}$}{#2}}}

\def\fpage#1{\begingroup
\voffset=.3in
\thispagestyle{empty}\begin{table}[b]\centerline{\footnotesize #1}
        \end{table}\endgroup}

\def\runninghead#1#2{\pagestyle{myheadings}
\markboth{{\eightit{\quad #1}}\hfill}{\hfill{\eightit{#2\quad}}}}

\font\tenbf=cmbx10
\font\tenit=cmti10
\font\tenit=cmti10
\font\bfit=cmbxti10 at 10pt
 1
 1
 1

\font\ninerm=cmr9

\font\eightrm=cmr8
\font\eightit=cmti8

\def\sqr#1#2{{\vcenter{\vbox{\hrule height.#2pt
          \hbox{\vrule width.#2pt height#1pt \kern#1pt
           \vrule width.#2pt}
           \hrule height.#2pt}}}}
\def\square{\mathchoice\sqr68\sqr68\sqr{4.2}6\sqr{3}6}

\def\qed{\hbox{${\vcenter{\vbox{                          %HOLLOW SQUARE
   \hrule height 0.4pt\hbox{\vrule width 0.4pt height 6pt
   \kern5pt\vrule width 0.4pt}\hrule height 0.4pt}}}$}}

\runninghead{Local Modes, Local Vacuum, Local Bogoljubov
Transformations and
$\ldots$} {Local Modes, Local Vacuum, Local Bogoljubov
Transformations and
$\ldots$}

\begin{document}
\normalsize\textlineskip
{\thispagestyle{empty}
\setcounter{page}{1}

\renewcommand{\thefootnote}{\fnsymbol{footnote}} %use symbolic footnote

%print out the publisher copyright heading
\copyrightheading{Vol. 0, No. 0 (1993) 000--000}

\vspace*{0.88truein}

\fpage{1}
\centerline{\bf LOCAL MODES, LOCAL VACUUM,}
\vspace*{0.035truein}
\centerline{\bf LOCAL BOGOLJUBOV COEFFICIENTS AND}
\vspace*{0.035truein}
\centerline{\bf  THE RENORMALISED STRESS TENSOR}
\vspace{0.37truein}
\centerline{\footnotesize S.MASSAR\footnote{Boursier I.I.S.N., e-mail:
smassar@ulb.ac.be}} \vspace*{0.015truein}
\centerline{\footnotesize\it Service de Physique Th\'eorique, Universit\'e
Libre de Bruxelles}
\baselineskip=10pt
\centerline{\footnotesize\it Campus Plaine, C.P. 225, Bd du Triomphe,
B-1050 Brussels, Belgium}
\vspace{0.225truein}
\publisher{(received date)}{(revised date)}

\vspace*{0.21truein}
\abstracts{\noindent Local modes and local particles are defined at
any point in curved space time as those that most resemble
Minkowsky modes at that point. It is shown that the renormalised
stress tensor is the difference of energy between the physical vacuum
and that defined by these local modes. }{}{}

\vspace*{-3pt}\textlineskip
\section{Introduction}
\noindent
Some difficulties have arisen in giving a
precise meaning to the notion of particles in curved
space time\cite{HAW}\cite{HAJ1}\cite{HAJ2}
(for a review on the subject of quantum fields in
curved space-time see \cite{BD}). These are due to the
curvature of space, which
creates particles and thus generates an uncertainty in their number.

In this article  (which generalises some
previous ideas of ours\cite{MPB2} ) we try to
clarify this issue. We proceed via the construction at an arbitrary point
in space time $\cal P$ of a set of local modes chosen so as to most
closely resemble the Minkowskian ones. These
modes then describe the local particles.
The justification of our procedure stems from the fact that the value of
the renormalised energy momentum tensor at $\cal P$ is the energy
momentum of the local particles at $\cal P$ (up to an anomalous
term depending only on the geometry at $\cal P$).

Since the local modes are those which an inertial
observer at $\cal P$ would most naturally choose as a basis for
quantisation, it appears that the gravitational field
strongly resembles inertial particle detectors, being
insensitive to the infinite energy density in the local vacuum and
responding only to the excitations thereof.

The number of particles present at the space time point $\cal P$ under
consideration is given a precise meaning by introducing the (local)
Bogoljubov coefficients between the local modes and the physical modes.
In this way, contact is made with the usual manner of exhibiting particle
creation through the Bogoljubov coefficients
between asymptotic modes  (as in
Hawking's original derivation of
black hole radiance\cite{HAW}), since in flat regions of
space time the local modes and the asymptotic modes coincide.

\textheight=7.8truein
\setcounter{footnote}{0}
\renewcommand{\thefootnote}{\alph{footnote}}

\section{Two Different Vacuum States}
\noindent
As a preparatory exercise we first recall how two different
quantisations of a scalar field are related by Bogoljubov coefficients.
Let $\phi$ be a real scalar field obeying the equation
\begin{equation}\square \phi + m^2 \phi + \xi R \phi =0 \label{I}
\end{equation}
and let $f_k(x)$ and $g_l(x)$ be two
complete orthonormalised sets of solutions of
(\ref{I}). The operator
$\phi$ may then be decomposed either as
\begin{equation}
\phi = \sum_k a_k f_k + a_k^+ f_k^* \label{II}
\end{equation}
or as
\begin{equation}\phi = \sum_l b_l g_l + b_l^+ g_l^* \label{III}
\end{equation}
The corresponding vacuum states $\vert O_f >$ and $\vert O_g >$ are
annihilated by the $a_k$ and $b_l$ respectively. The Bogoljubov
coefficients relating these two bases are given by
\begin{equation}\alpha_{k l} = <f_k,g_l> \qquad\qquad
\beta_{k l} = <f_k^*,g_l> \label{IV}
\end{equation}
where $<,>$ is the Klein-Gordon inner product. The $\alpha$ and $\beta$
coefficients obey unitary relations
\begin{equation}\alpha^2-\beta^2 =1
\qquad\qquad\alpha\beta^* - \alpha^*\beta =0
\label{V}
\end{equation}
The propagators $G_f(x,x^\prime)=<O_f\vert \phi(x) \phi(x^\prime)\vert
O_f>$ and $G_g(x,x^\prime)=<O_g\vert \phi(x) \phi(x^\prime)\vert
O_g>$ are singular at the coincidence point but their difference
$G_f-G_g$ is not. This is a consequence of Hadamard's theorem,\cite{HAD}
which states that Green's functions of Eq. (\ref{I}) always have the
same singular part. Similarly, the difference
of energy between the $f$ and
$g$ vacuum is finite and is given by
\begin{equation}<O_f\vert T_{\mu\nu}(x)\vert O_f> -
<O_g\vert T_{\mu\nu}(x)\vert O_g>
= \lim_{x^\prime \rightarrow x} \hat T_{\mu\nu}
\left( G_f(x,x^\prime)
- G_g(x,x^\prime) \right) \label{VI}
\end{equation}
where $\hat T_{\mu\nu}$ is a differential operator whose explicit form
we do not need here. We note for future use that the difference of
energy (\ref{VI}) is a conserved quantity.
This is classicaly true and results here
from the fact that $G_f - G_g$ obeys (\ref{I})
in both variables and is non
singular.\cite{BO}

The difference of the propagators can be expressed in terms of the $g$
modes and the Bogoljubov coefficients
(i.e. in terms of the number of $g$ particles
present in the $f$ vacuum):
\begin{equation}G_f(x,x^\prime) - G_g(x,x^\prime) =
\int \!dldl^\prime\!dk \ 2 {\cal R}e
\left \lbrack \beta_{lk}\beta^*_{l^\prime k}
g^*_l(x) g_{l^\prime}(x^\prime)
- \alpha_{lk}\beta^*_{l^\prime k}
g_l(x) g_{l^\prime}(x^\prime)\right \rbrack \label{VII} \end{equation}
 where we have used the unitarity relations (\ref{VI}).
Acting on (\ref{VII}) with $\hat
T_{\mu\nu}$ and taking the coincidence point limit
yields an expression for the difference
of energies as a function of the Bogoljubov coefficients.

\textheight=7.8truein
\setcounter{footnote}{0}
\renewcommand{\thefootnote}{\alph{footnote}}

\section{Local Modes, Local Vacuum, Local Bogoljubov Coefficients and the
Renormalised Stress Tensor}
\noindent
We now turn to the problem at hand, namely the definition of the
local quanta and the local vacuum. To this end choose a specific point
$\cal P$ in space time and build Riemann normal coordinates $y^\mu$
centred on $\cal P$ (i.e. the coordinate lines are the geodesics
passing through $\cal P$). In these coordinates the metric is given by
\begin{equation}g_{\mu\nu} = \eta_{\mu\nu} + {1\over 3}
R_{\mu\alpha\nu\beta} y^\alpha y^\beta + O(R^2) \label{VIII}
\end{equation}
and the wave equation (\ref{I}) takes the form
\begin{equation}\left(\eta_{\mu\nu}\partial_\mu\partial_\nu + m^2 +\xi R +
+ {1\over 3}R^\nu_\alpha y^\alpha \partial_\nu
- {1\over 3}R^{\mu\ \nu}_{\ \alpha\ \beta}
y^\alpha y^\beta\partial_\mu\partial_\nu
+ O(R^2)\right)
\phi =0 \label{IX}
\end{equation}
The local modes are the complete set of solutions of (\ref{IX}) that start
off as Minkowsky modes. They can be computed order by order in the
curvature:
\begin{eqnarray}\phi_{local\ k} =
{e^{i k_\mu y^\mu} \over \sqrt{ (2 \pi)^3 k_0}}\left(
1 +
{\xi R \over m^2}
-{i\over 3 m^2}R^\nu_\alpha y^\alpha k_\nu
-{i\over 3 m^2}R^{\mu\ \nu}_{\ \alpha\ \beta}
y^\alpha y^\beta k_\mu k_\nu \ +  \right.
\nonumber \\
\qquad\qquad\qquad\quad\left. +\
{4i\over 3 m^4}R^{\mu\ \nu\sigma}_{\ \alpha} y^\alpha k_\mu
k_\nu k_\sigma +{8\over 3 m^8}R^{\mu\nu\sigma\rho}
k_\mu k_\nu k_\sigma k_\rho + O(R^2)
\right) \label{X}
\end{eqnarray}
Having defined the local modes we proceed exactly
as in the preceding section: the
states of interest are the local vacuum $\vert O_{local}(\cal P) >$ and the
physical state (taken for simplicity to be a vacuum state $\vert
O_{in}>$ ).

The propagator in the local vacuum is
\begin{equation}G_{local}(y^\mu,y^{\prime \mu})=
<O_{local}({\cal P})\vert \phi(y^\mu)\phi(y^{\prime\mu})\vert
O_{local}({\cal P})>=\int\!d^3\!k \ \phi_{local\ k}(y^\mu)
\phi_{local\ k}^*(y^{\prime\mu}) \label{XI}
\end{equation}
Note that the local vacuum and the local propagator
are parametrised by the space time
point $\cal P$ and will in general change as $\cal P$ changes.

The renormalised stress tensor is the difference of energy
between the physical state and the local vacuum
\begin{eqnarray}<T_{\mu\nu}({\cal P})>_{ren}=
\lim_{y^\mu \rightarrow 0 \atop y^{\prime\mu} \rightarrow 0}
\hat T_{\mu\nu} \left[\
<O_{in}\vert \right. \ \phi(y^\mu)
\phi(y^{\prime\mu})\vert O_{in}> -  \nonumber \\
  \qquad\qquad\qquad\qquad\qquad\qquad
\left. - <O_{local}({\cal P})\vert \phi(y^\mu)\phi(y^{\prime\mu})\vert
O_{local}({\cal P})>\ \right] \label{XII}
\end{eqnarray}

Eq. (\ref{XII}) coincides with the usual expressions for $<T_{\mu\nu}>_{ren}$
found in the literature. This can be readily seen by considering the
function $G_{local}(y^\mu,0)$ given by (\ref{XI}) with $y^\prime=0$, which
coincides with the adiabatic propagator
defined by Bunch and Parker\cite{BP} (both
functions are solutions of the same equation and are built as a series in
the curvature starting with the Minkowskian form, therefore they are
identical). Bunch and Parker's\cite{BP} proof of the equivalence of their
renormalisation procedure and the renormalisation of the DeWitt-Schwinger
effective action then carries through unaltered.

As written  above, the renormalised stress
tensor (\ref{XII}) is not conserved. This
arises on taking the divergence of the
right hand side of (\ref{XII}) from the
space-time dependence of the local vacuum
$\vert O_{local}(\cal P) >$. This situation
must be compared with the
difference in energy of two fixed states (\ref{VI}),
which is conserved. A
conserved renormalised stress tensor may be
recovered by adding to (\ref{XII}) a
term depending only on $m^2$, $\xi$ and the
geometry at $\cal P$ (since the
subtraction in (\ref{XII}) depends only on
these quantities).\cite{BO} \cite{BF} In
the conformally coupled case this term, the
anomaly, depends only on the geometry at
$\cal P$  and restores energy conservation at the expense of conformal
invariance, i.e. the stress tensor now has a trace.

Finally, Bogoljubov coefficients between the modes describing the physical
state and the local modes can be constructed and the renormalised stress
tensor expressed in terms of these coefficients, that is,
in terms of the number of local
particles present in the physical state.

\textheight=7.8truein
\setcounter{footnote}{0}
\renewcommand{\thefootnote}{\alph{footnote}}

\section{One Dimensional Case and Black Hole Evaporation}
\noindent
The above formalism has been applied \cite{MPB2}
to the situation of a massless
field in $1+1$ dimensions. In that case the series
for the local modes (\ref{XI}) may be
summed in closed form. After some manipulations the
renormalised stress tensor can be
expressed as a sum over poles. This result has in turn been used in
the effective one-dimensional problem describing black hole
evaporation.\cite{UNR} It then  appears that in
the part of the stress tensor describing the
outgoing flux the poles tend to their thermal
value at infinity but differ from it for finite
distances, thereby illustrating how the
Hawking radiation builds up to its asymptotic value.

\nonumsection{Acknowledgements}
\noindent
The author wishes to thank R. Brout, R. Parentani and Ph. Spindel for
important discussions and R. Parentani for a careful reading of the manuscript.

\nonumsection{References}


\noindent

\begin{thebibliography}{000}

\bibitem{HAW}
S.W. Hawking, Commun. math. Phys. {\bf 43},
199 (1975).

\bibitem{HAJ1}
P. H\'{a}j\'{\i}\v{c}ek, Nuovo Cimento {\bf 33} B, 597 (1976).

\bibitem{HAJ2}
P. H\'{a}j\'{\i}\v{c}ek, Phys. Rev. D {\bf 15}, 2757 (1977).

\bibitem{BD}
N.D. Birrel and P.C.W. Davies, Quantum Fields in Curved Space,
Cambridge University Press , Cambridge, England (1982).

\bibitem{MPB2}
S. Massar, R. Parentani, R. Brout, Brussels
preprint ULB-TH-01/93, hep-th 9303147

\bibitem{HAD}
J. Hadamard, Lectures on Cauchy's Problem in
Linear  Partial Differential Equations, Yale
University Press, New Haven, 1923

\bibitem{BO}
M.R. Brown and A.C. Ottewill, Phys. Rev. D {\bf 34}, 1776 (1986)

\bibitem{BP}
T.S. Bunch and L. Parker, Phys. Rev. D {\bf 20}, 2499 (1979).

\bibitem{UNR}
W. G. Unruh, Phys. Rev. D {\bf 14}, 870 (1976)

\bibitem{BF}
D. Bernard and A. Folacci, Phys. Rev. D {\bf 34}, 2286 (1986)

\end{thebibliography}
\end{document}